\RequirePackage{fix-cm}
\documentclass[onecolumn]{svjour3}
\makeatletter
\let\cl@chapter\relax
\makeatother

\journalname{Eur. Phys. J. C}
\usepackage[pdfborder={0 0 0},citecolor=blue,linkcolor=blue,colorlinks]{hyperref}
\usepackage[utf8]{inputenc}
\usepackage[T1]{fontenc}    
\usepackage{subcaption}
\usepackage[format=hang,labelfont=bf,hypcap=true]{caption}
\usepackage{xspace}
\usepackage{graphicx} 
\usepackage[usenames,dvipsnames]{xcolor}

\usepackage[capitalize,nameinlink,compress]{cleveref}

\usepackage{todonotes}
\usepackage{listings}
\definecolor{listingsbg}{rgb}{0.65,0.95,0.85}
\definecolor{yamlkey}{rgb}{0.00,0.50,0.00}      
\definecolor{yamlstr}{rgb}{0.73,0.13,0.13}      
\definecolor{yamlcomment}{rgb}{0.24,0.48,0.48}  
\newcommand\YAMLkeystyle{\color{yamlkey}\bfseries}
\newcommand\YAMLvaluestyle{\color{black}\mdseries}
\newcommand\YAMLcolonstyle{\color{black}}
\lstdefinelanguage{yaml}{
  sensitive=true,
  keywords={true,false,null,yes,no},
  keywordstyle=\color{yamlstr}\bfseries,
  comment=[l]{\#},
  commentstyle=\color{yamlcomment}\itshape,
  morestring=[b]',
  morestring=[b]",
  stringstyle=\color{yamlstr},
  moredelim=**[il][\YAMLcolonstyle{:}\YAMLvaluestyle]{:},
  moredelim=[l][\YAMLvaluestyle]{-\ },
  basicstyle=\YAMLkeystyle,
}
\lstdefinestyle{yamlstyle}{
  language=yaml,
  backgroundcolor=\color{listingsbg},
  basicstyle=\scriptsize\ttfamily\YAMLkeystyle,
  breaklines=true,
  columns=fullflexible,
  keepspaces=true,
  showstringspaces=false,
}
\usepackage{booktabs}

\usepackage{tocloft}
\usepackage{float}            
\newfloat{listing}{htbp}{lol} 
\floatname{listing}{Example}
\crefname{listing}{Example}{Examples}

\newcommand{\epem}{\ensuremath{\mathrm{e}^+\mathrm{e}^-}\xspace}

\newcommand{\keyhep}{\textsf{Key4hep}\xspace}
\newcommand{\GeneratorsConfig}{\textsf{k4GeneratorsConfig}\xspace}

\newcommand{\edmhep}{\textsc{EDM4hep}\xspace}
\newcommand{\yamlinput}{\texttt{YamlInput}\xspace}
\newcommand{\parametersets}{\texttt{ParameterSets}\xspace}
\newcommand{\parameters}{\texttt{Parameters}\xspace}
\newcommand{\parameter}{\texttt{Parameter}\xspace}
\newcommand{\process}{\texttt{Process}\xspace}
\newcommand{\generators}{\texttt{Generators}\xspace}
\newcommand{\generator}{\texttt{Generator}\xspace}
\newcommand{\generatorBase}{\texttt{GeneratorBase}\xspace}
\newcommand{\generatorProcDB}{\texttt{GeneratorProcDB}\xspace}
\newcommand{\procDBBase}{\texttt{ProcDBBase}\xspace}

\newcommand{\babayaga}{\textsc{Babayaga}\xspace}
\newcommand{\kkmc}{\textsc{KKMCee}\xspace}
\newcommand{\pythia}{P\protect\scalebox{0.8}{YTHIA}\xspace}
\newcommand{\herwig}{H\protect\scalebox{0.8}{ERWIG}\xspace}
\newcommand{\madgraph}{M\protect\scalebox{0.8}{AD}G\protect\scalebox{0.8}{RAPH}\xspace}
\newcommand{\mg}{\madgraph}

\newcommand{\whizard}{W\protect\scalebox{0.8}{HIZARD}\xspace}
\newcommand{\rivet}{R\protect\scalebox{0.8}{IVET}\xspace}
\newcommand{\sherpa}{S\protect\scalebox{0.8}{HERPA}\xspace}
\newcommand{\HepMC}{H\protect\scalebox{0.8}{EP}MC\xspace}

\newcommand{\finalize}{\texttt{finalize}\xspace}

\newcommand{\python}{\texttt{Python}\xspace}
\newcommand{\cpp}{\texttt{C++}\xspace}
\newcommand{\importlib}{\texttt{importlib}\xspace}
\newcommand{\str}{\texttt{str}\xspace}
\newcommand{\dict}{\texttt{dict}\xspace}
\newcommand{\sqrtS}{\ensuremath{\sqrt{s}}\xspace}

\begin{document}

\title{Configuration and Benchmarking of \epem Processes with \GeneratorsConfig}
\author{Alan Price        \and
        Dirk Zerwas 
}

\institute{A. Price \at
              Jagiellonian University,ul. prof. Stanis\l{}awa \L{}ojasiewicza 11, \\
              30-348 Kraków, Poland \\
              \email{Alan.Price@uj.edu.pl}           
           \and
           D. Zerwas \at
              DMLab, Deutsches Elektronen-Synchrotron DESY, \\
              CNRS/IN2P3, Notkestr. 85, Hamburg, Germany \\
              \email{dirk.zerwas@in2p3.fr}
}

\date{Received: 23 March 2026 / Accepted: 23 June 2026}
\maketitle
\begin{abstract}
    The next generation of electron–positron colliders will require unprecedented precision in both theory and experiment. Sophisticated software frameworks are essential to evaluate detector concepts, optimize designs, and simulating physical processes. In this context, Monte Carlo (MC) event generators play a central role, enabling realistic simulations of Standard Model processes and providing the basis for physics studies. However, technical consistency across different generators remains critical, particularly in domains where agreement is expected. To address this need, we present \GeneratorsConfig, a Python-based package that automates the benchmarking process for MC generators. The tool translates universal physics inputs into generator-specific configurations, ensuring consistency, reproducibility, and reduced human error. Its modular design allows for straightforward integration of additional generators, supports batch processing, and provides compatibility with the Key4hep software stack.
\end{abstract}

\section{Introduction}
\noindent
    An electron-positron Higgs/top/electroweak factory is considered to be a leading candidate for 
    the next collider project.
	Whether this machine is ultimately realized as a circular collider such as FCC/CEPC~\cite{FCC:2018evy,FCC:2025uan,FCC:2025lpp,CEPCStudyGroup:2018ghi,CEPCStudyGroup:2023quu} or a linear collider LCF/ILC/CLIC~\cite{LinearColliderVision:2025hlt,Behnke:2013xla,Aicheler:2012bya}, it will demand substantial advancements across all areas of particle physics—from theoretical predictions to detector technologies. 
	Such large scale experiments, realized for example at the (HL)-LHC such as ATLAS~\cite{ATLAS:2008xda,ATLAS:2010arf} and CMS~\cite{CMS:2008xjf}, require a sophisticated software chain to study the physical detector responses of a wide range of physical processes. 

\noindent
	The European Committee for Future Accelerators (ECFA) organized a series of workshops on  Higgs, electroweak, and top-quark factories, which culminated in the final report~\cite{Altmann:2025feg}, highlighted not only the physics potential of future electron-positron colliders but also identified the critical role of software frameworks in enabling these experiments~\cite{HEPSoftwareFoundation:2025irs}. The study emphasizes that sophisticated simulation and analysis tools are essential to evaluate detector concepts, optimize designs, and benchmark physics processes in a reproducible way. In particular, such frameworks must bridge diverse Monte Carlo (MC) event generators, supporting consistent and reliable comparisons across different implementations. The report underscores the need for modular, extensible software that can handle varying collider designs—whether linear or circular—while maintaining precision, flexibility, and reproducibility, which are vital for planning and validating future high-energy experiments.

\noindent
	One such avenue of development is the improvement of MC event generators~\cite{Alwall:2014hca,Kilian:2007gr,Sherpa:2024mfk,Bahr:2008pv,Bierlich:2022pfr,Jadach:2022mbe,CarloniCalame:2000pz,Campbell:2022qmc,Bellm:2015jjp,Frixione:2021zdp,Carloni:2019ejp,Balossini:2006wc,Krauss:2022ajk} which will play a central role in this ecosystem and will be essential to the success of any future collider program. Those based on developments during the LEP era, maintained and some updated from fortran to \cpp, are often specialized for a narrow range of Standard Model (SM) processes or based on the implementation of a separately calculated matrix element. Modern MC generators leverage automated algorithms that enable them to handle a much wider array of processes with enhanced flexibility and precision.
	Despite differences in design and implementation, MC generators are expected to agree in certain foundational areas~\cite{Ballestrero:2000ur,Grunewald:2000ju,Andersen:2024czj,Aliberti:2024fpq,Altarelli:1996ww,CERN:1995gra,Jadach:2000zp,Korneeva:2024oho,Karneyeu:2013aha}, such as leading-order (LO) cross-sections. 

\noindent
The Python package, \GeneratorsConfig automates the benchmarking process for MC event generation. The central philosophy behind this package is that the physics we wish to simulate for our tests should remain independent of the specific MC generator being used. With this goal in mind, we designed the tool to accept universal physics inputs, which are then systematically translated into generator-specific input formats required for simulation.
The physics information provided to the tool can range from simple details like the collision energy to more complex specifications, such as beam types, particle distributions, and the particular processes to simulate. This abstraction ensures that users can define the physics of their simulation straightforwardly and consistently without needing to account for the idiosyncrasies of different MC generators. Once the input is provided, the package generates runcards or configuration files tailored to the syntax, parameters, and requirements of the specific generators.

\noindent
\GeneratorsConfig  ensures that the results presented further on in this section are robust and repeatable. By concentrating on areas where discrepancies are unlikely, we can provide a valuable diagnostic insights for both users and developers of MC generators. Such comparisons help uncover technical issues, validate algorithmic implementations, and ensure that these tools meet the stringent precision demands of future collider experiments.

\noindent
\GeneratorsConfig supports a variety of MC generators relevant for \epem colliders, making it a versatile solution for diverse simulation needs. It not only handles the physics parameters but also configures essential simulation settings, such as random seed values, number of events, and technical details required by the generators. This automation ensures compatibility across generators. 
The tool’s modular architecture also enables seamless extensibility. New generators can be added by defining their respective templates or translation rules.

\noindent
The alpha release of \GeneratorsConfig is currently available both as a standalone package and as an integrated component of the \keyhep~\cite{Key4hep:2022xly} software stack, which is used by all future collider projects. As a standalone tool, it focuses on generating the required runcards for various Monte Carlo generators based on user-provided physics inputs. Furthermore it provides scripts to automatically execute the corresponding generators within \keyhep.
Working in \keyhep the package can launch the event generation step and provide comparisons process by process 
between generators. Additionally \GeneratorsConfig offers the capability to convert all generated Monte Carlo events into \edmhep~\cite{Gaede:2021izq,Gaede:2022leb}.

\noindent
The tests for the \GeneratorsConfig package, using github actions, run a Continuous Integration (CI) pipeline based on a \keyhep software stack providing the workflows. This integration ensures that the software is consistently validated with every update or change. By leveraging the CI infrastructure, the package undergoes a series of automated tests that check its functionality, compatibility, and performance across different environments. These tests include validating the generation of runcards, checking the conversion of physics information into generator-specific inputs, and ensuring that the results are reproducible and consistent across different Monte Carlo generators. The CI system runs these tests on multiple platforms, guaranteeing that any issues are detected early in the development process. This automated testing process greatly enhances the reliability and stability of the \GeneratorsConfig package, making it easier for developers and users to integrate the tool into their workflows with confidence.


\section{Implementation}
\label{sec:implementation}
\noindent
The software suite is implemented in \python, with process configurations specified in a YAML-formatted input file. This file defines the process to be simulated, the generators to be used, and other relevant settings. For each unique combination of process, generator, and centre-of-mass energy, a separate output directory is created. Each directory contains generator-specific input files derived from the common YAML configuration, along with a shell script to execute the generator within the \keyhep environment. The shell scripts can be easily modified to use local installations of specific generator packages.

\noindent
This functionality is complemented by being able to run CI checks from the \python script as well as the actual
event generation step as well as producing a summary for all processes, center--of--mass energies and generators
requested. Care was taken to be able to run the steps independently as well as jointly in a single execution.
The former allows for splitting the full task into parallel tasks.  

\subsection{Defining the input}
\label{sec:yamlInput}

\begin{table}[htb]
    \centering
    \renewcommand{\arraystretch}{1.2}
    \begin{tabular}{@{}lp{10cm}@{}}
        \toprule
        \textbf{Key} & \textbf{Description} \\
        \midrule
        \texttt{Generators}            & List of strings specifying which generators to use \\
        \texttt{Model}                 & String indicating the model name \\
        \texttt{SqrtS}                 & Float value for center-of-mass energy in GeV\\
        \texttt{Events}               & Integer number of events to generate \\
        \texttt{RandomSeed}           & Integer seed for random number generation \\
        \texttt{OutDir}               & String specifying output directory \\
        \texttt{OutputFormat}         & String: one of \texttt{edm4hep}, \texttt{hepmc3}, or \texttt{lhe} (default: \texttt{edm4hep}) \\
        \texttt{EventMode}            & String: \texttt{weighted} or \texttt{unweighted} \\
        \texttt{NLO}                  & Integer flag for next-to-leading order (NLO) mode, off: 0 (default), on : 1\\
        \texttt{EWParamDevThreshold}            & Float: Maximum allowed relative deviation between EW LO parameters (default 0.001) \\
        \midrule
        \texttt{PolarisationDensity}  & List of two integers: \texttt{[-1 or 1, 1 or -1]} (default: \texttt{[-1, 1]}) \\
        \texttt{PolarisationFraction} & List of two floats in the range [0.0, 1.0] \\
        \texttt{Beamstrahlung}        & String specifying beamstrahlung model or file \\
        \midrule
        \texttt{Processes}            & Dictionary keyed by user-defined \textit{ProcessName}. Each entry includes: \\
                                      & \hspace{1em} \texttt{Final: [PDGIDs]}, \texttt{RandomSeed}, \texttt{ISRMode: 1/0} \\
        \midrule
        \texttt{ParticleData}         & Dictionary keyed by \textit{PDGID}, each with \texttt{mass} and \texttt{width} values \\
        \midrule
        \texttt{Selectors}            & Optional nested dictionary (keyed by \texttt{Process: \textit{ProcessName}}) with selection criteria: \\
                                      & \hspace{1em} \texttt{PT, ET, Energy, Rapidity, Eta, Theta, Mass, Angle,} \\
                                      & \hspace{1em} \texttt{DeltaEta, DeltaRapidity, DeltaPhi, DeltaR, Flavour} \\
        \midrule
        \texttt{Analysis}             & Analysis tools to use: \texttt{Tools: [key4hep, rivet]} \\
                                      & Rivet analysis to use: \texttt{RivetAnalysis: [MC\_XS, MC\_ZINC,\dots]} \\
                                      & Rivet path: \texttt{RivetPath: /path/to/analysis} \\
        \bottomrule
    \end{tabular}
    \caption{Main keys of the \yamlinput configuration file along with their types and descriptions, maintained with examples at  \url{https://key4hep.github.io/k4GeneratorsConfig}. \textit{ProcessName} is a user-defined string. All keys are internally converted to lowercase.}
    \label{tab:yamlinput}
\end{table}

\noindent
The main keys of the \yamlinput are listed in \Cref{tab:yamlinput}, which are case insensitive, and
are stored internally after conversion to lower case. The general configuration keys define
the generators to be configured, the model to be used, the $\sqrt{s}$, the number of events to be generated,
the random number seed and the top level output directory. The EventMode can be defined as weighted or unweighted
as the user wishes. \texttt{Final}  is the list of final state 
particles using the particle data group integer scheme (PDGID). Initial state radiation
can be switched off or on with \texttt{ISRmode}.
As a \yamlinput may have several processes defined, keys such as \texttt{RandomSeed} can be redefined for 
each process, overwriting the global definition.
\texttt{ParticleData} is a structured list of PDGIDs~\cite{ParticleDataGroup:2024cfk}. Each PDGID has a \texttt{mass} and/or \texttt{width} assignment
for the particle. Each process is assigned a unique identifier string,\texttt{ProcessName},
 which is used to prevent file overwriting and to facilitate organized storage of all associated outputs.

\noindent
\texttt{Selectors} allows the user to define phasespace restrictions, such as cut on the polar angle of 
final state particle, either globally for all processes or on a process-by-process basis. If no \texttt{ProcessName} is listed, the 
criteria are applied to all processes defined in the \yamlinput. 
The \texttt{Flavour} field is a list of PDGIDs to
which the criteria should be applied, eg., transverse momentum/energy or angle. For 
composite variables like the invariant mass or angular differences, two PDGIDs have to be provided.
For each variable a \texttt{MIN} and a \texttt{MAX} field define the numerical values to be applied. 
To ensure consistency the selector keys of \Cref{tab:yamlinput} are cross-checked with
the selector keys implemented for each generator, triggering an exception on failure.

\noindent
The luminosity spectrum of electron-positron collisions is affected by beam--beam interactions and such
effects can be modelled by MC generators. The spectrum does depend on the specific accelerator which
can be defined with the key \texttt{Beamstrahlung}. Currently, the spectrum is provided by 
CIRCE~\cite{Ohl:1996fi}, as is the case for \whizard,  or provided by a generator-specific implementation, as is the case for \mg.

\noindent
Since linear colliders are expected to offer the possibility of beam polarization, the software includes options to specify the polarization of the incoming beams. This is done using the discrete parameter \texttt{PolarisationDensity}, which accepts a list of integers representing the polarization states of the incoming particles. Additionally, the \texttt{PolarisationFraction} parameter defines the average polarization of the two beams. Currently, we have limited this to the longitudinal polarization of the incoming beams, as this is supported by more than one generator. This can be extended to include transverse polarization if we becomes more common across generators.

\noindent
The \texttt{Analysis} key triggers the addition of the execution of an analysis to the shell script 
after the event generation has been run. Either an automated analysis in the \keyhep framework or a \rivet 
analysis can be run. This will be further discussed in~\Cref{sec:analysis}.
\noindent
Finally, it is worth noting that not all settings available in one generator are available in all. To circumvent this issue, we have also implemented a generator-specific setting that can be used for that generator alone. It can be added to the \yamlinput in the generic form \texttt{Generator:Key:Value}, where \texttt{Key} will be the \generator specific setting, which could potentially consist of additional subkeys. These settings will take priority over the main keys in the \yamlinput and could potentially lead to incorrect set-ups for keys that have not been tested~\footnote{A complete list of tested keys will be maintained at \url{https://key4hep.github.io/k4GeneratorsConfig}}.
\subsection{Model Parameters}
\label{sec:input}

\noindent
To ensure a consistent definition of model parameters, a global set is implemented in the \parameters module, where each parameter is defined as a \parameter object with a fixed default value. While most Standard Model input parameters—such as the Fermi constant—have remained stable, the framework allows users to override these defaults when needed. Parameter sets can be defined directly within the main configuration file or provided in a separate YAML-formatted file. Each set is associated with a unique tag, which can be specified via the command line. Additionally, a custom parameter file can also be selected at runtime using a command-line option.

\noindent
One consideration for the input to MC generators is the choice of input parameters, with specific numerical values, which must be chosen in a manner that ensures the results are gauge invariant and logically consistent, both in the chosen input parameters and in the corresponding derived parameters. In particular, one has some freedom in the choice of electroweak input schemes as described in Section~5.1 of~\cite{Denner:2019vbn}. 

\noindent
In our package, the electroweak scheme with the Fermi constant ($G_F$)  and the
masses of the electroweak gauge bosons ($m_\textrm{W}$ and $m_\textrm{Z}$) is used. 
Most of the generators, e.g., \madgraph, \sherpa and \whizard, support this scheme while others, e.g.,  \kkmc, \herwig and \pythia need the electroweak mixing angle, with \herwig also requiring the 
inverse of the electromagnetic coupling constant at zero momentum transfer. As a consequence either a scheme translation has to be performed on the fly, possibly having to include radiative corrections to translate the parameters into
effective coupling constants. Or the parameter set is extended to include the derived 
parameters as well. The latter solution was chosen for this package.

\begin{table}[htb]
    \centering
    \renewcommand{\arraystretch}{1.2}
    \begin{tabular}{@{}lll@{}}
        \toprule{}
        \textbf{Parameter} & \textbf{Name} & \textbf{Default Value} \\
        \midrule
        GFermi             & $G_F$                                 & $1.16638 \cdot 10^{-5}\ \textrm{GeV}^{-2}$ \\
        MW                 & $m_\textrm{W}$                        & 80.379 GeV \\
        MZ                 & $m_\textrm{Z}$                        & 91.1876 GeV \\
        \midrule
        alphaSMZ           & $\alpha_S(m_\textrm{Z})$             & 0.1184 \\
        \midrule
        alphaEMMZM1        & $\alpha_{\textrm{EM}}(m_\textrm{Z})^{-1}$             & 127.9 \\
        alphaEMMZ          & $\alpha_{\textrm{EM}}(m_\textrm{Z})$                  & $1/\alpha_{\textrm{EM}}(m_\textrm{Z})^{-1}$ \\
        alphaEMLOM1        & $\alpha_{\textrm{EM}}(m_\textrm{Z})^{-1}_\textrm{LO}$ & 132.507 \\
        alphaEMEWLO        & $\alpha_{\textrm{EM}}(m_\textrm{Z})_\textrm{LO}$      & $1/\alpha_{\textrm{EM}}(m_\textrm{Z})^{-1}_\textrm{LO}$ \\
        alphaEMM1          & $\alpha_{\textrm{EM}}(0)^{-1}$                         & 137.035999139 \\
        alphaEM            & $\alpha_{\textrm{EM}}(0)$                              & $1/\alpha_{\textrm{EM}}(0)^{-1}$ \\
        sin2theta          & $\sin^2\vartheta$                                     & 0.23155 \\
        sin2thetaEff       & $\sin^2\vartheta_{\textrm{eff}}$                      & 0.23155 \\
        sin2thetaLO        & $\sin^2\vartheta_{\textrm{LO}}$                       & 0.22276773 \\
        \midrule
        VEV                & $v$                                                  & 246 GeV \\
        \midrule
        MH                 & $m_\textrm{H}$                                       & 125 GeV \\
        MT                 & $m_\textrm{t}$                                       & 172 GeV \\
        ymt                & $y_\textrm{t}$                                       & $m_\textrm{t}/v$ \\
        MB                 & $m_\textrm{b}$                                       & 4.7 GeV \\
        ymb                & $y_\textrm{b}$                                       & $m_\textrm{b}/v$ \\
        \midrule
        WZ                 & $\Gamma_\textrm{Z}$                                  & 2.4952 GeV \\
        WW                 & $\Gamma_\textrm{W}$                                  & 2.085 GeV \\
        WH                 & $\Gamma_\textrm{H}$                                  & 0.00407 GeV \\
        WT                 & $\Gamma_\textrm{t}$                                  & 1.50833649 GeV \\
        \bottomrule
    \end{tabular}
    \caption{Parameters defined in the \texttt{\parametersets} and their default values.}
    \label{tab:ParameterSet}
\end{table}

\noindent
The full list of allowed parameters is provided in \Cref{tab:ParameterSet}. To maintain internal consistency, certain parameters are treated as derived quantities such as the electromagnetic coupling constant as its inverse is customarily used. The same approach applies to Yukawa couplings, which are computed from the corresponding quark masses and the vacuum expectation value(VEV). The latter defined as $2\cdot m_\textrm{Z}/\sqrt{g_1^2+g_2^2}\sim 246~\textrm{GeV}$ with $g_1$ and $g_2$ the coupling constants
associated to the $U(1)_Y$ and $SU(2)_L$ groups of the Standard Model.

\noindent
Any parameter values associated with a specific tag in \parametersets will override the corresponding defaults. To ensure that the results are consistent and gauge-invariant, the user must specify a coherent input
parameter scheme based on the phenomenology of their study. A cross check is performed at leading
order to check the consistency of $G_F$, $m_\textrm{W}$ and $m_\textrm{Z}$ with
the electroweak 
mixing angle ($\sin^2\vartheta_{\textrm{LO}}$), the electromagnetic coupling constant ($\alpha_{\textrm{EM}}(m_\textrm{Z})_\textrm{LO}$), its inverse ($\alpha_{\textrm{EM}}(m_\textrm{Z})_\textrm{LO}^{-1}$) and the vacuum expectation value ($v$) after the reading of the \parametersets using the tree-level relations. Any relative
deviation larger than the \texttt{EWParamDevThreshold} parameter is reported as a warning with a default setting of permil (0.001).

\noindent
For the electromagnetic coupling constant $\alpha_\textrm{EM}^{-1}$, three definitions are supported: the value at $Q = 0$, the value at $Q = m_\textrm{Z}$, and the leading-order (LO) value derived from the standard electroweak input scheme.
Additionally, the parameter set includes both the weak mixing angle and the effective weak mixing angle. The strong coupling constant $\alpha_s$ is defined at the scale $Q = m_\textrm{Z}$.
In addition to the masses of the electroweak bosons, the fermion masses of the top quark and the bottom quark
are part of the parameter set as well as their Yukawa couplings. 
The Yukawa couplings for the default values are derived from the ratio of the masses and the vacuum expectation value. If an exclusive final state, e.g., of the Higgs boson, is requested, some generators require
the specification of the particle width. Therefore the widths of the W, Z bosons, of the Standard Model Higgs boson and of the top quark
are initialized. 


\subsection{Command Line Options}
\label{sec:commandline}

\noindent
The package is shipped with multiple setup scripts, depending on the users shell environment, which defines an alias \texttt{\GeneratorsConfig} serving as executable. The four types of processing modes as described in \Cref{sec:implementation}
\begin{enumerate}
    \item \texttt{-{}-make}: creates generator datacards and shell scripts
    \item \texttt{-{}-check}: compares the created cards and scripts to reference files
    \item \texttt{-{}-generate}: runs the event generation with the scripts and datacards
    \item \texttt{-{}-summary}: produces a summary of all processes, generators and center--of--mass energies
\end{enumerate}
are shown in \Cref{tab:commandLine}.
As a shortcut \texttt{-{}-all} runs the creation, generation and summary steps in a single execution.

\begin{table}[htb]
    \centering
    \begin{tabular}{@{}lllp{5.5cm}@{}}
        \toprule
        \textbf{Key} & \textbf{Type} & \textbf{Default} & \textbf{Function} \\
        \midrule
        \texttt{-{}-make}     & flag   & false & make the generator datacards from the yaml files \\
        \texttt{-{}-check}            & flag   & false & check the generator datacards with respect to the reference \\
        \texttt{-{}-generate}         & flag   & false & run the event generation\\
        \texttt{-{}-summary}          & flag   & false & compare the results of the event generation process by process, produce summary output in outputDir \\
        \texttt{-{}-all}              & flag   & false & activates \texttt{-{}-make -{}-generate -{}-summary} \\
        \texttt{-{}-outputDir} & string & Run-Cards & location of the generator datacards and shell scrippts, active for all modes\\             
        \midrule
        \texttt{-{}-yaml}              & string           & ---     & Path to the YAML input files/directories \\
        \texttt{-{}-nevts}             & integer          & -1      & Number of events to generate \\
        \texttt{-{}-seed}              & integer          & 4711    & Random number seed \\
        \texttt{-{}-sqrts}             & list of floats/file   & ---     & List of center-of-mass energies\\
        \texttt{-{}-parameterTag}      & string           & latest  & Name of the parameter tag to use \\
        \texttt{-{}-parameterTagFile}  & string           & ---     & Path to a YAML file with parameter sets \\
        \texttt{-{}-key4hepVersion}     & string (date)    & ---     & Specific Key4HEP release in YYYY-MM-DD format \\
        \texttt{-{}-key4hepUseNightlies} & flag            & false   & Use nightly Key4HEP builds instead of stable releases \\
        \midrule
        \texttt{-{}-refDir}           & string & k4GeneratorsConfig/test/ref-results & path to the reference files \\ 
        \midrule
        \texttt{-{}-generator}         & string   & All & generator to be run, active for \texttt{-{}-check} and \texttt{-{}-generate} \\
        \bottomrule
    \end{tabular}
    \caption{Command-line options with their types, functions, and default values. The flags in the first section
    steer the type of processing to be performed. The output directory is valid for all types of processing. Checking and generating can be restricted to a single generator. The second section applies only to the creation of datacards.}
    \label{tab:commandLine}
\end{table}

\noindent
Several other command line options are implemented and listed in \Cref{tab:commandLine}. 
Most are applicable to a particular processing step with the exception of \texttt{-{}-outputDir}, which specifies
the location of the generator datacards and shell scripts as well the \texttt{-{}-generator} option which allows
to restrict the checking and the event generation step to a single generator.

\noindent
When running the \texttt{-{}-make} option, a \yamlinput file containing the definitions described in \Cref{sec:input} has to be provided as an argument to the 
\texttt{-{}-yaml} command line option. This option can take lists as well as directories as input.

\noindent
The optional arguments provide a mechanism to override values in the \yamlinput such as the 
number of events to be generated and random number seed to be used for the first process to be configured. The center of mass 
energies can be given as a list and/or in a file. In the file the energies should be specified again as a list with the key \texttt{ecms: [240., 500.]}.
Since a list of \yamlinput can be given as argument, the same set of center of mass energies can be run for different
inputs in a single call.


\noindent
The version of \keyhep to be run in the event generation step can be configured as well.
The default is the latest release of \keyhep for which no specification is needed. A specific version can be requested by specifying
the date of the version in the format YYYY-MM-DD. A consistency check of the requested date is performed. 
The presence of the \texttt{-{}-key4hepUseNightlies} option switches the configuration to the use of the \keyhep nightlies. This
option can be used together with the option to define the date of the release to choose a particular nightly instead of the latest one. 
This allows the user to easily recreate previous results using the exact same software framework.

\subsection{Process Configuration}
\label{sec:config}
\noindent 
The \process module fills the data structures of the given processes. The beam particles and their properties are identified as well as the particles of the final state. Furthermore the process, i.e., the initial and final state unsigned
PDGIDs, are concatenated in a label. This label is used later by the generator process database to 
add process dependent settings to the datacard.
The creation of input cards and execution scripts for running generators within the \keyhep environment is managed by the \generators module. For each \generator, a dedicated Python module translates the requested processes into the generator-specific input format. Upon receiving a \yamlinput for a process, the \generators module loads the corresponding generator module dynamically. The datacard required to run the generator is first built in memory, then written to the directory \textit{OutDir}/\generator/\textit{ProcessName}[\_ecms]. The energy in MeV is appended to the directory if multiple centre-of-mass energies are specified. Generator-specific modules and classes are dynamically imported via Python’s \importlib. To add a new \generator, it is sufficient to implement a corresponding module and class—no changes to the \generators steering module are required. The \generator string and the class name must match exactly, including case sensitivity. 
\noindent
An inheritance scheme is used in the implementation as each \generator inherits from the \generatorBase class. 
\generatorBase provides several services to each \generator. During the initialization of the
class, the process specific and generator specific
settings are stored in a public data member. 
The names of the files with the \generator specific extension and the \yamlinput specified output directory are built
and stored in public data members. While in general only two files, a datacard for the \generator and the shell script
to run in the \keyhep environment are necessary, a third optional file structure is available. The optional file
is used, e.g., by the \pythia class to store the requested generator cuts on the process.
For \sherpa, \whizard, \madgraph, \herwig and \babayaga the datacards are generated on the fly. 
For \kkmc, a template file is loaded and only a limited number of predefined settings can be changed via
string replacement.

\noindent
For the file contents of the datacard, script and optional file, data encapsulation is implemented
to avoid accidental overwriting. Access methods are provided to 
allow adding \str content, either as an addition or as a replacement. Furthermore, accessor methods for the removal of options and a full reset of the contents are provided as well.
Since much of the work is performed in \generatorBase, \generator specific information is made accessible
through required functions. This can be the way to set the particle mass or width or the format of the
option to set a parameter.

\noindent
For the \keyhep script the setup of the environment is provided and configurable as described in \Cref{sec:commandline}.
Additionally, a separate structure to define analyses on the generated Monte Carlo data is available. 
The content of this structure is appended to the \keyhep script at the \finalize step before the output to disk.   
While many steps in the datacard generation are done dynamically, some settings specific to processes and/or
particle properties, may be provided as a de facto library. An example for the former is the definition
of the process. \sherpa, \whizard and \madgraph are based on the 
definition of the initial and final state only, which corresponds to the choice made for the \yamlinput format.
\pythia and \herwig on the other hand defines a process as the type of mediator in addition to both the initial and final states encoded in
a key word.

\begin{figure}
    \centering
    \includegraphics[width=\textwidth]{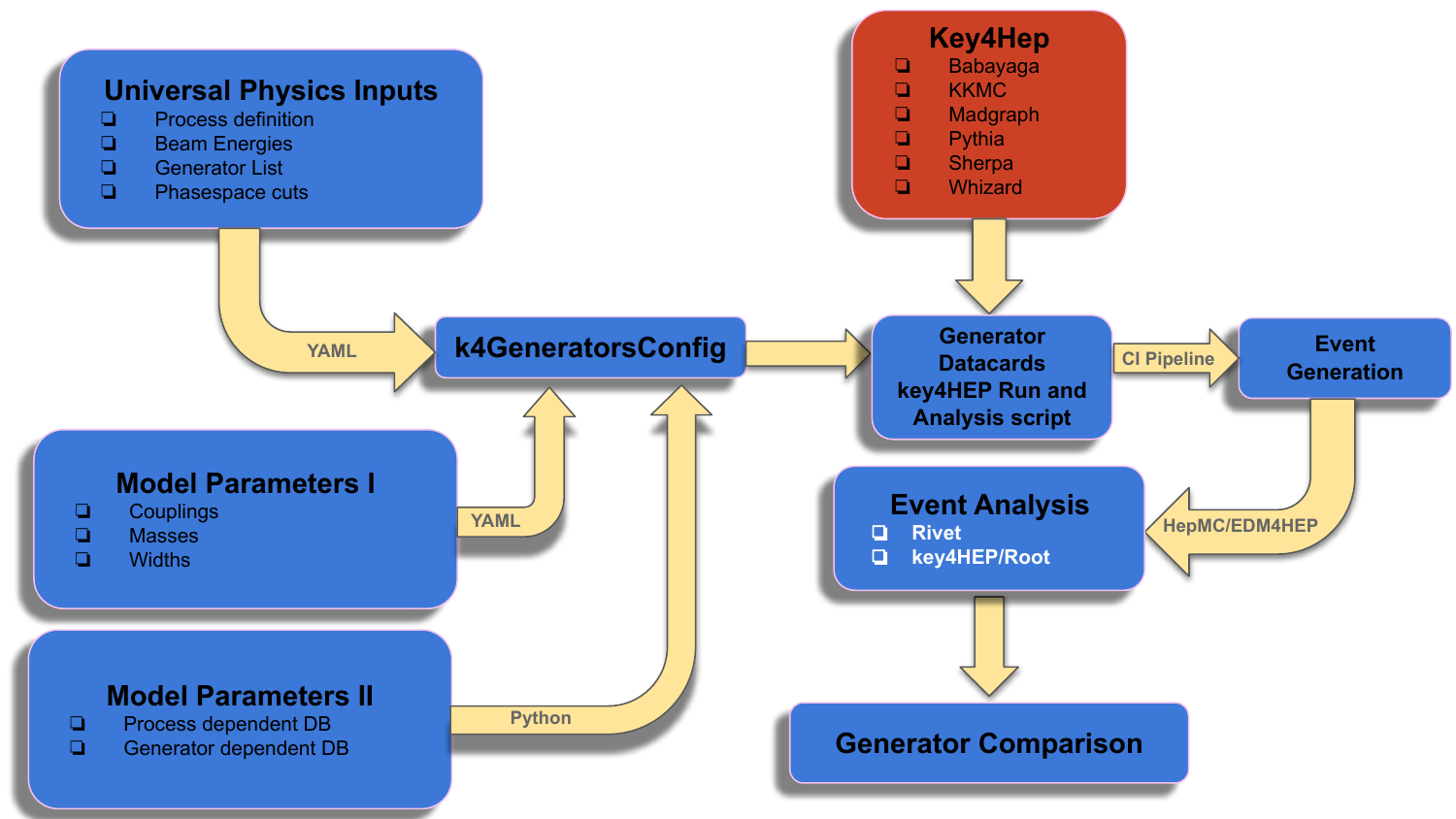}
    \caption{Workflow of the \GeneratorsConfig framework: universal physics inputs and model parameters are combined into a unified YAML configuration, which generates input cards for multiple event generators via Key4Hep. Event generation and analysis outputs are compared across generators.}
    \label{fig:workflow}
\end{figure}

\noindent
A generator specific process database \generatorProcDB is loaded for each \generator if implemented.
Each \generatorProcDB inherits from the \procDBBase class. 
\procDBBase provides a data structure to store, access and remove the standard option specific to the 
generator and to the process being configured. For simplicity all options are stored as simple \python \dict.
Providing a \generatorProcDB is not mandatory, as in certain cases, e.g., \kkmc, inputs cannot be added dynamically. 
Conflicts may arise when the \yamlinput specifies properties which are also defined by one or 
more \generatorProcDB libraries. In this case the \yamlinput supersedes the \generatorProcDB settings.
Particle properties may be specified as part of the \parametersets, e.g. the electroweak gauge bosons, 
in the \yamlinput and in the \generatorProcDB. The hierarchy is implemented as \parametersets settings
taking precedent over \generatorProcDB and \yamlinput superseding \parametersets.
In~\Cref{fig:workflow}, we show a workflow diagram of how the \GeneratorsConfig tool is used.

\noindent
Consistency checks are performed at the generator level. Beamstrahlung can only be applied
for \whizard and \madgraph. For \madgraph the chosen closest match is chosen and printed out.
Polarisation is available only for \madgraph, \sherpa and \whizard.
For two of the LEP--era generators strict checks are applied.
\kkmc is restricted to fermion-anti--fermion pair production (excluding top pairs). 
In the \generatorProcDB the final state must be implemented explicitly for \pythia and \herwig.
Furthermore for \babayaga only electron beams in the initial state and pairs of electrons, muons and photons in the final state are allowed. 

\noindent
While the package was developed for \epem{} collisions, care was taken to be compatible for
a muon collider as well. In the \generatorProcDB the process specific settings are chosen with
a tag as fully defined process (initial and final state), where the initial state can be either electron-positron
pairs or muon-anti-muon pairs. Higgs--Bremsstrahlung at a muon collider 
has been implemented in the CI examples.

\subsection{Example \yamlinput}
\label{sec:example}

\begin{listing}
\centering
\begin{minipage}{0.5\linewidth}
\begin{lstlisting}[style=yamlstyle]
Generators:
  - Herwig
  - KKMC
  - Madgraph
  - Pythia
  - Sherpa
  - Whizard

Events: 10000
OutputFormat: edm4hep
OutDir: Run-Cards
EventMode: unweighted
SqrtS: 91.2
Model: SM
ISRMode: 0

Processes:
  Muon91.2:
     Final: [13, -13]

  Muon250:
     Final: [13, -13]
     SqrtS: 250
     ISRMode: 1

Selectors:
  Process:
    Muon250:
      PT:
        Max: 175
        Min: 20
        Flavour: [-13,13]
    Muon91.2:
      PT:
        Max: 45.6
        Min: 20
        Flavour: [-13,13]

Sherpa:
  Run:
    EW_SCHEME: 3

ParticleData:
  23:
    Mass: 91.1876
    Width: 2.4952

Analysis:
  Tools: [key4HEP]

\end{lstlisting}
\end{minipage}
\caption{An example of a \yamlinput for \GeneratorsConfig.}
\label{eg:proc}
\end{listing}
\noindent
An example of the \yamlinput is shown in~\Cref{eg:proc}.
The first step is to define the \generators to be run and the number of events.
Furthermore the output directory, the output event format and the type of events to be generated is
configured.
The \sqrtS, and the initial state radiation setting defined here are valid for all processes.
The process of interest is then defined here, $\epem \rightarrow \mu^+\mu^-$. For the moment
we assume the initial state beams will be \epem, this just leaves the final state to be defined.
Here we have actually defined two process with the same final state, $\mu^+\mu^-$,
but at different \sqrtS. For the second process we have turned on initial state radiation, superseeding for this process the global setting for all processes. 
In the lower part of~\Cref{eg:proc} phase space restrictions are implemented for the two processes. 
A minimal and maximal transverse momentum is defined to be applied to 
the listed particle flavours. Some particle properties are set, super-seeding the default values.
Specific settings for generators can be added as shown in the example for \sherpa.
Finally, the automated \keyhep analysis is run after event generation.

\section{Analysis}
\label{sec:analysis}
\noindent
Fully differential MC generators give us the ability to analyse the complicated 
final state kinematics for the physical states simulated within the generator.
Such distributions can then be compared to data, where we assume the data has
been corrected for detector effects, to test the validity 
of our calculations. It is also a crucial requirement that generators,
which on paper are equivalent at a given level of precision, also 
agree with each at a differential level. For this reason, we have included
an analysis framework within \keyhep, that allows the user to specify 
differential distributions in which we compare. We currently support 
an interface to both \rivet~\cite{Buckley:2010ar,Bierlich:2019rhm,Bierlich:2024vqo} and  Root packages~\cite{Brun:1997pa}. 

\subsection{Rivet}
\label{sec:rivet}

\noindent
To analyse the events that are generated in our workflow, we have provided
an in-house interface to the \rivet software package
which is a commonly used analysis toolkit for the validation
of Monte Carlo event generators, either against experimental data or 
as a comparison tool between multiple MC generators.
Provided the events are in format that \rivet supports, such a \HepMC, the event record is 
directly read from the file, which is automatically 
created. 
The only additional information required is the 
name of \rivet analyses, and the corresponding paths, that are to be applied to the final state event
record as shown in~\Cref{tab:yamlinput}.

\subsection{\keyhep}
\label{sec:keyhep}

\noindent
Another functionality is the Monte Carlo truth analysis in the \keyhep environment. If requested
via the \yamlinput file, the analysis is run after the event generation step.
Several histograms are prepared for the final state particles specified in the \yamlinput.
For all final state particles the $\cos\vartheta$ in the lab frame is provided. 
Furthermore, for all two-particle combinations of the final states, the invariant mass,
transverse momentum and longitudinal momentum of the pair is binned. 
In the simplest case of a two-body final state with initial state radiation the distributions characterise 
the partonic system in the lab frame. 

\noindent
If several processes for several $\sqrtS$ are run, a process summary is generated, triggered by the command line
option \texttt{-{}-summary}. This summary
consists of the cross section and the differential distributions
provided as figures with all available generators overlaid (with a fixed colour and symbol code) as well 
as the deviation from the average. While the cross section extraction from the \keyhep files tests
the writing of the meta data of the generator as well as their conversion from \HepMC to \edmhep, the differential distributions test the event record.

\subsection{Results}

\begin{figure}[htb]
\centering
\includegraphics[width=0.33\linewidth]{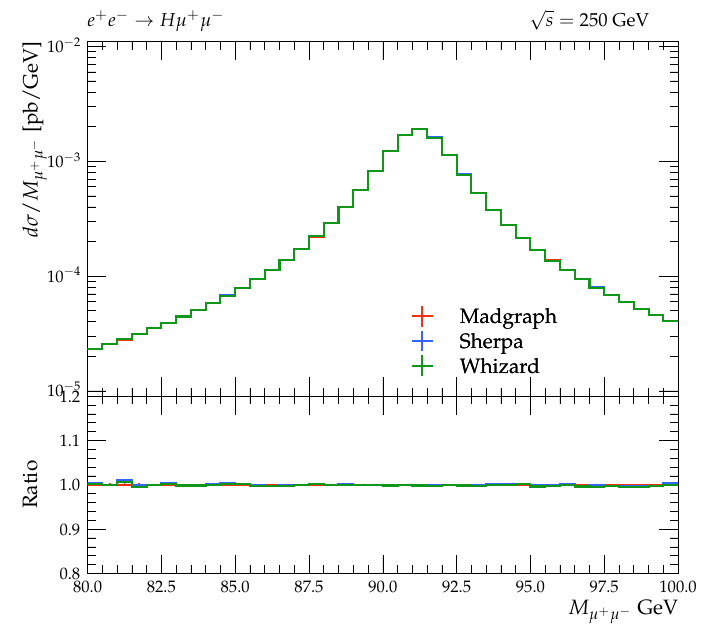}
\includegraphics[width=0.33\linewidth]{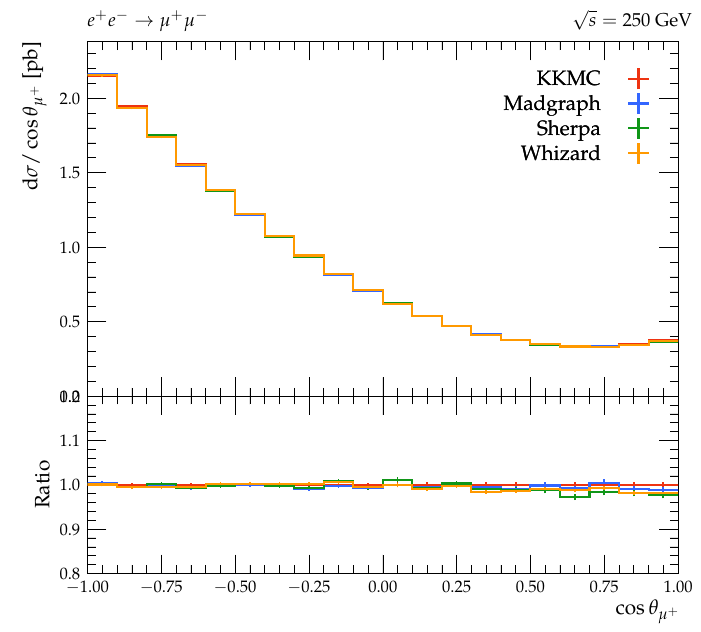}
\includegraphics[width=0.33\linewidth]{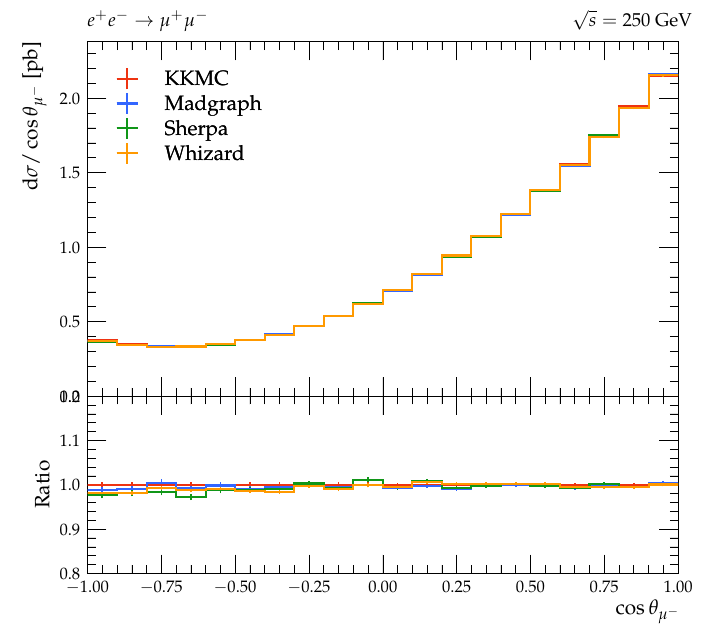}
\caption{Example output from the \rivet interface of the \GeneratorsConfig 
framework for two processes: Higgsstrahlung production and the corresponding 
difermion background. (Top left) Invariant mass distribution of the final lepton 
pairs in Higgsstrahlung production. (Top right, bottom) Polar angle distributions 
in the difermion background process. These plots are intended to demonstrate the 
functionality of the \rivet interface and the automated comparison workflow; 
high-statistics quantitative comparisons between generators are foreseen as part 
of future work.}
\label{plot:rivet}
\end{figure}

\noindent

\noindent
In~\Cref{plot:rivet}, we show example output from the \rivet interface of the 
\GeneratorsConfig framework. The primary purpose of these plots is to demonstrate 
that the interface is fully operational and to illustrate how results are 
presented to the user: multiple generators are overlaid on the same distribution, 
with bin-by-bin deviations flagged automatically when significant differences 
are observed. We show distributions for two processes, namely Higgsstrahlung and 
the corresponding difermion background, both generated from a single \yamlinput 
file. The Higgsstrahlung process is only available in a subset of the generators 
considered, while the difermion background is more widely supported, for example 
in \kkmc. High-statistics quantitative comparisons between generators, and 
validation against experimental data where available, are foreseen as part of 
future work and go beyond the scope of the present technical description of the 
framework.

\begin{figure}[htb]
  \centering
  \includegraphics[width=0.33\linewidth]{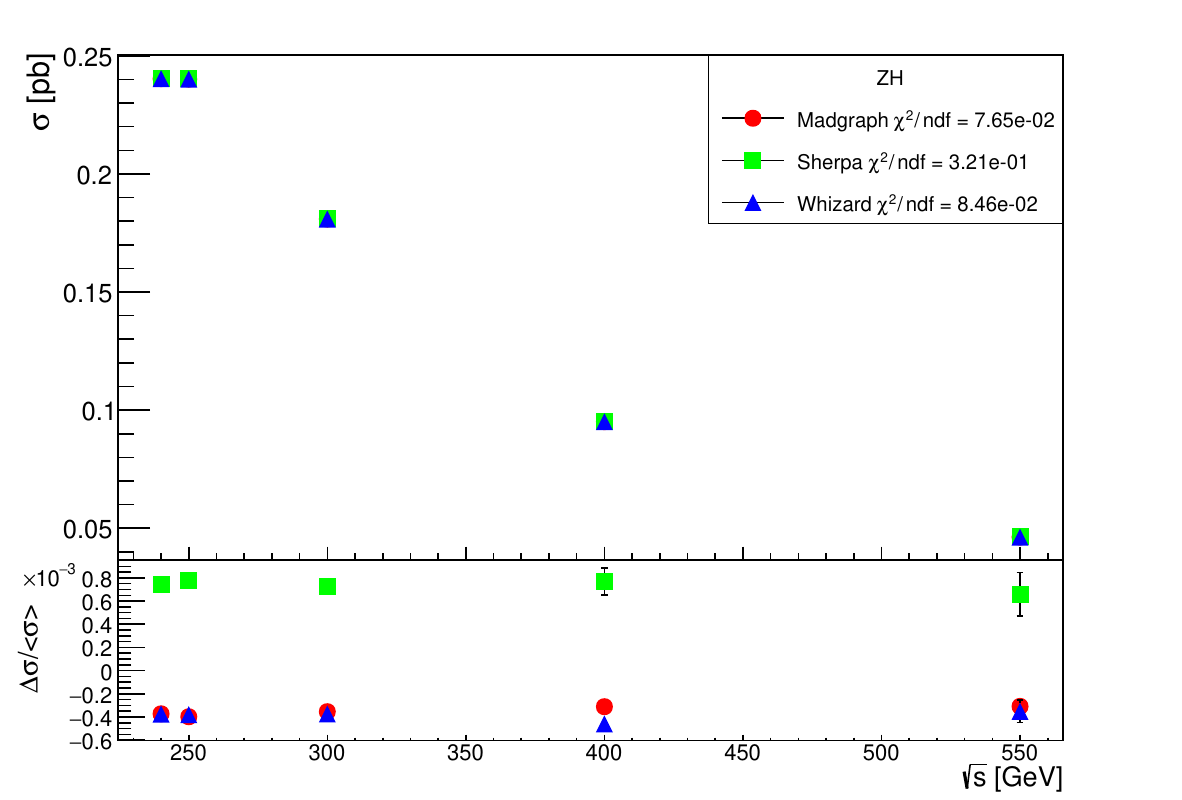}
  \includegraphics[width=0.33\linewidth]{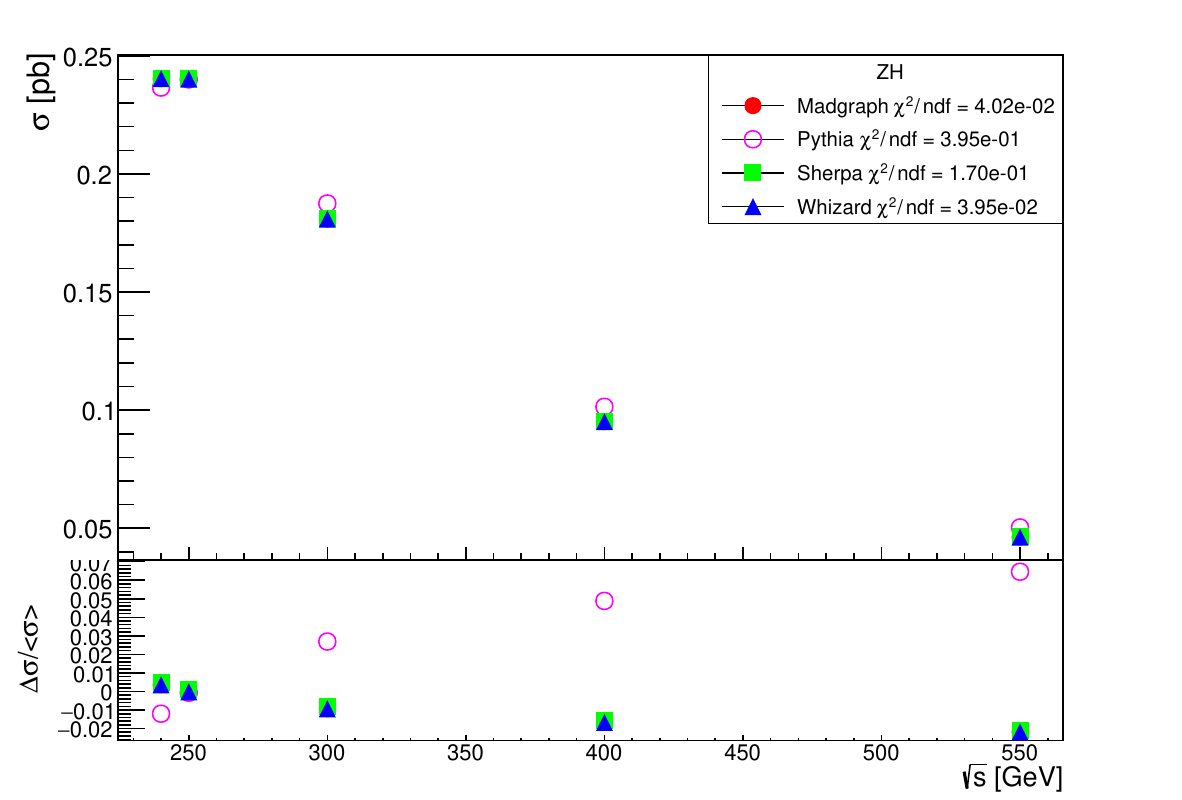}
  \includegraphics[width=0.33\linewidth]{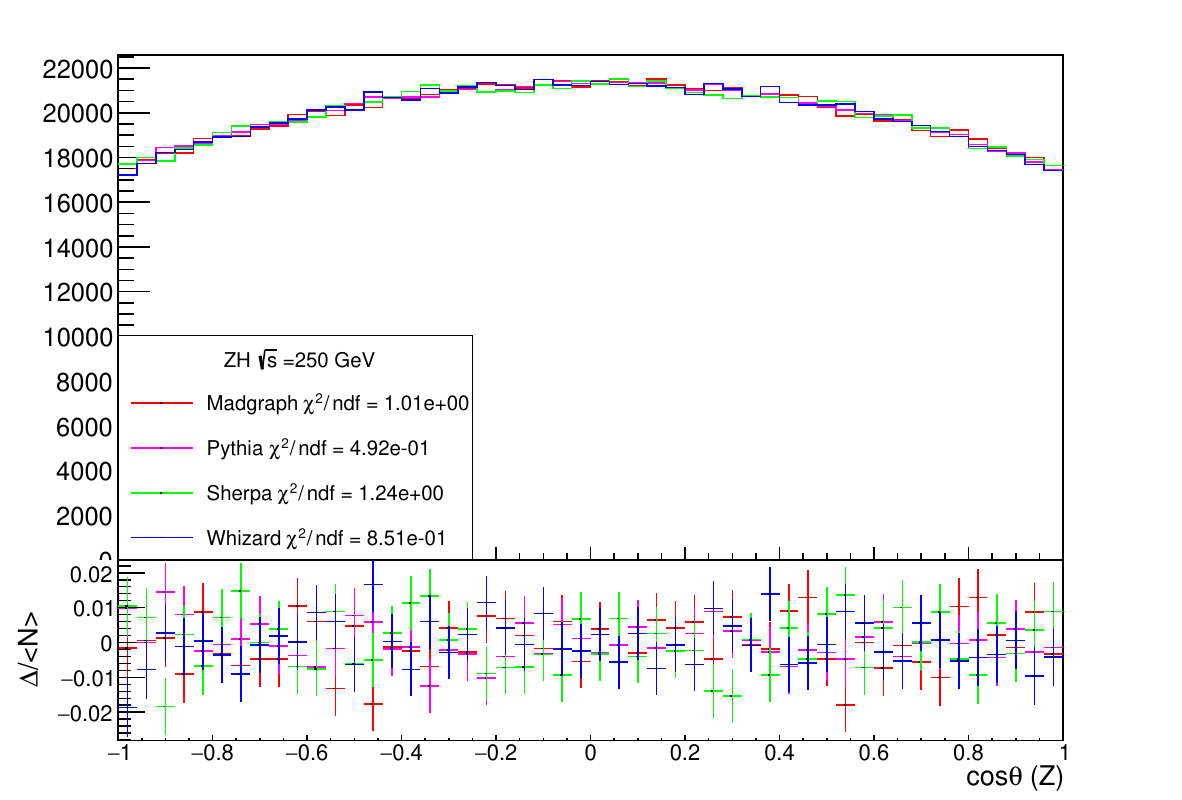}
  \caption{(Top left) The result of LO calculation of ZH production is shown as function of $\sqrt{s}$ for three generators in the top panel. In the bottom panel the deviation from the average cross section of shown for all generators. (Top right) \pythia is show in addition.
  (Bottom) The distribution of the polar angle of the Z~boson at $\sqrt{s}=250\textrm{GeV}$ is shown as well as the deviation from the average per bin is shown in the bottom panel for the four generators.}
  \label{fig:key4hepZHxsec}
\end{figure}

\noindent
For the analysis with \keyhep, the results are shown in
in \Cref{fig:key4hepZHxsec}. The cross section is shown as function of $\sqrt{s}$ for several generators in the top left and right panels. The deviation 
from the average cross section is shown separately for each generator. As a measure of compatibility, a $\chi^2/\textrm{ndf}$ is provided as well. As the comparison is with respect to the average, ndf is the number 
of energy points. As not all generators provide a reliable error estimate, the error on the cross section
average is used in the $\chi^2$ calculation.

\noindent
\sherpa, \whizard and \madgraph are in excellent agreement, better than one-permil with \whizard and \madgraph agreeing at
the level of a tenth of permil. This is also reflected in the $\chi^2/\textrm{ndf}$ value which is slightly higher 
for \sherpa.
The prediction by \pythia is slightly off at the percent level as shown in the right panel, reflected also in the 
higher $\chi^2/\textrm{ndf}$ value.
This is understood as \pythia requires the input of the effective electroweak mixing angle and a value within the errors on the experimental measurements. Translation of the EW scheme to the \pythia input leads to a rejection of the calculated input
which is required to be higher as \pythia expects a $\sin^2\vartheta$ including radiative corrections. Therefore a true leading order comparison cannot include \pythia~\cite{Bierlich:2022pfr}.

\noindent
In the bottom panel of \Cref{fig:key4hepZHxsec} the distribution of the polar angle of the Z~boson is shown for a center of
mass energy of 250~GeV. Each bin is compared to the average of the generators. All generators are compatible with the average with the statistical 
error. Using a $\chi^2/\textrm{ndf}$ measure the values are similar and below unity, where the ndf is the number 
of bins and the error is the statistical error of the distribution for each generator.

\noindent
Using the \GeneratorsConfig package several problems have been identified and 
corrected~\footnote{For example, see 
\url{https://key4hep.github.io/k4GeneratorsConfig/found-issues.html}}.

\begin{itemize}

    \item \textbf{Incomplete generator configuration in \keyhep:} One generator 
    was found to behave differently across platforms within the CI pipeline, 
    which was traced back to an incomplete configuration of the generator 
    environment in \keyhep. The automated, platform-independent nature of the 
    CI tests was essential in exposing this issue, which would not have been 
    apparent from single-platform local testing.

    \item \textbf{Missing cross section output:} Several generators were found 
    not to write cross section information to the output metadata. This was 
    identified through the \keyhep summary step, which aggregates cross sections 
    across generators and $\sqrt{s}$ values. The absence of this information 
    would silently compromise any downstream reweighting or normalisation 
    procedure that relies on the generator cross section.

    \item \textbf{Rounding error leading to event rejection:} A non-trivial 
    numerical issue was identified in which insufficient precision in the 
    internal representation of a physical parameter, such as a particle mass 
    or width, caused events to be rejected at the matrix element or phase 
    space level. The effect manifested as a systematic deviation in the 
    predicted cross section relative to other generators across a range of 
    $\sqrt{s}$ values, rather than as an obvious numerical failure. This 
    demonstrates that cross-generator comparison at the level of integrated cross 
    sections provides sensitivity to subtle numerical precision issues that would 
    be invisible in single-generator validation.

    \item \textbf{ISR and boost inconsistencies:} An inconsistency in the 
    treatment of initial-state radiation was identified through cross section 
    comparisons: events were being rejected at the parton shower or hadronisation 
    stage due to a misconfiguration of the ISR or lab-frame boost settings. This 
    issue is particularly insidious because the effect is invisible at the level 
    of the hard-process cross section alone, the generator appeared to run 
    successfully, but produced physically inconsistent samples. Only by comparing 
    differential distributions and event counts systematically across generators 
    was the problem identified.

\end{itemize}

\noindent
These examples illustrate that the \GeneratorsConfig framework provides 
sensitivity to qualitatively distinct classes of problem: platform-dependent 
configuration errors, missing metadata, numerical precision issues, and 
physics-level misconfigurations. In each case, the automated and systematic 
nature of the cross-generator comparisons was essential to identifying the issue 
quickly and unambiguously.


\section{Conclusions}
\noindent
In this work, we have presented the \GeneratorsConfig package, a robust  d automated framework for validating event generators in the context of next-generation collider experiments. By ensuring the consistent production of input configurations, together with integrated tools for event generation and validation within the \keyhep environment, the package enables systematic cross-comparisons across different generators and software releases.

\noindent
As part of ongoing developments, we plan to extend the automated tests within the \keyhep CI pipeline. These include the re-generation of runcards, validation of physics accuracy, and verification of software compatibility with updated generators. 
We also plan to expand our validation suite to include more state-of-the-art predictions that are provided by the generators and systematically include these tests within the \keyhep CI. Such continuous integration ensures that changes or improvements to the generators are seamlessly incorporated without introducing regressions, thereby strengthening the reliability of large-scale simulations for future colliders.

\noindent
Automated process-by-process comparisons across multiple generators have already been implemented in the CI workflow. Initial tests revealed deviations in certain generators, which were traced back to implementation issues, thereby demonstrating the value of inter-generator validation as a complement to the dedicated efforts of individual generator groups. Future work will extend these comparisons to a broader set of processes, further enhancing the robustness of event generation for forthcoming Higgs factories.
Further work will also include comparisons between different versions of a generator. This 
can be used to pinpoint the origin of deviations. The implementation is strongly related to the architecture of the software stack.

\noindent
While the current implementation focuses on leading-order comparisons, the 
framework is designed with higher-order extensions in mind. Aligning 
next-to-leading order (NLO) and beyond predictions across generators presents 
several additional challenges. At NLO and beyond, generators may differ in their 
choice of infrared subtraction scheme (e.g.\ Catani--Seymour versus 
FKS), renormalisation and factorisation scale settings, and the treatment of 
parton shower matching and merging (e.g.\ MC@NLO versus POWHEG). These 
differences can produce scheme-dependent variations that are not indicative of 
errors but must nevertheless be understood and documented before meaningful 
cross-generator comparisons can be made. A further complication arises from 
the fact that not all generators support the same set of higher-order processes, 
making process-by-process coverage at NLO inherently asymmetric. Despite these 
challenges, the \GeneratorsConfig framework provides a natural foundation for 
such comparisons: the automated runcard generation and CI integration already 
ensure reproducibility at LO, and extending this to NLO requires primarily the 
standardisation of input parameters and the adoption of a common 
analysis level via \textsc{Rivet}.

\noindent
The precision demands of the next generation of Higgs factories will place unprecedented requirements on the supporting software ecosystem, with long-term stability being critical to experimental success. While event generator developers conduct detailed validations of their codes, the \GeneratorsConfig package provides an independent and complementary validation strategy. By enabling automated cross-checks against alternative generators and earlier releases, it ensures that users within the \keyhep framework can rely on physically consistent samples across diverse generators. This early detection of potential mismodelling within the simulation workflow offers a significant safeguard against inefficiencies and unnecessary computational expenditure, thereby supporting the long-term sustainability of precision collider physics simulations.

\section*{Acknowledgements}
\noindent
The work was started in Working Group \begin{it}Physics Analysis Tools\end{it} of the ECFA Study on Higgs/Top/Electroweak
Factories (2020--2025). We are grateful for the support provided by Patrizia Azzi and Fulvio Piccinini and the discussions
during the Working Group and Plenary meetings. The work would not have been possible without the \keyhep team, in particular
Thomas Madlener and Juan Miguel Carceller as well as Frank Gaede, Andr\'e Sailer and Gerardo Ganis. 
On the generator authors side we are grateful for many discussions with J\"urgen Reuter, Olivier Mattelaer, Marco Zaro and Stefano Frixione and others from the KKMC, Pythia, and Sherpa Monte Carlo groups. We gratefully acknowledge the feedback and comments on this manuscript provided by Enrico Bothmann, Fulvio Piccinini, Tomasz Procter, Thomas Madlener and Frank Gaede.
\noindent
The work of A.P. was supported by the OpenMAPP project via National Science Centre, Poland under CHIST-ERA programme
(grant No. NCN 2022/04/Y/ST2/00186) and by the Priority Research Area Digiworld under the program ‘Excellence Initiative – Research University’ at the Jagiellonian University in Kraków and the Marie Sk\l{}odowska-Curie grant agreement No 945422.
A.P gratefully acknowledges the Polish high-performance computing infrastructure PLGrid (HPC Centers: CI TASK, ACK Cyfronet AGH) for providing computer facilities and support within computational grant no. PLG/2024/017287.

\bibliographystyle{amsunsrt_mod}
\bibliography{journal.bib}

\end{document}